\documentclass[preprints,article,accept,moreauthors,10pt,a4paper]{mdpi} 
\firstpage{1} 
\articlenumber{x}
\doinum{10.3390/------}
\pubvolume{xx}
\pubyear{2017}
\copyrightyear{2017}
\externaleditor{Academic Editor: name}
\history{Received: date; Accepted: date; Published: date}

\usepackage{amsmath, amsthm, amssymb, amsfonts}

\Title{Hamiltonian dynamics of doubly-foliable space-times}

\Author{Cec\'\i lia Gergely$^{1}$, Zolt\'{a}n Keresztes$^{2}$ and L\'{a}szl\'{o} \'{A}rp\'{a}d Gergely$^{3,}$*}

\AuthorNames{Cec\'\i lia Gergely, Zolt\'{a}n Keresztes and L\'{a}szl\'{o} \'{A}rp\'{a}d Gergely}

\address[1]{%
$^{1}$ \quad Institute of Physics, University of Szeged, Hungary; lawrencesterne92@gmail.com\\
$^{2}$ \quad Institute of Physics, University of Szeged, Hungary; zkeresztes.zk@gmail.com\\
$^{3}$ \quad Institute of Physics, University of Szeged, Hungary; laszlo.a.gergely@gmail.com}

\corres{Correspondence: laszlo.a.gergely@gmail.com; Tel.: +36702020800}

\abstract{The 2+1+1 decomposition of space-time is useful in monitoring the temporal evolution of gravitational perturbations/waves in space-times with a spatial direction singled-out by symmetries. Such an approach based on a perpendicular double foliation has been employed in the framework of dark matter and dark energy motivated scalar-tensor gravitational theories for the discussion of the odd sector perturbations of spherically symmetric gravity. For the even sector however the perpendicularity has to be suppressed in order to allow for suitable gauge freedom, recovering the 10th metric variable. The 2+1+1 decomposition of the Einstein-Hilbert action leads to the identification of the canonical pairs, the Hamiltonian and momentum constraints. Hamiltonian dynamics is then derived via Poisson brackets.}

\keyword{space-time foliation; extrinsic curvature; normal fundamental form and scalar; symmetries; geometrodynamics}

\begin{document}

\section{Introduction}

In the curved space-time of general relativity gravitational waves propagate
with the speed of light (the velocity limit), correcting the Newtonian
description of gravity. Whenever a reference system is chosen, time needs to
be singled out. 
In the 3+1 decomposition of space-time, known as the
Arnowitt-Deser-Misner (ADM) formalism of gravity \cite{ADM}, the constant
time 3-surfaces form a foliation. While the time parameter is constant on
each hypersurface, it changes monotonically from one hypersurface to the
other. The role of the 4-dimensional metric $\widetilde{g}_{ab}$ (with$\ 10$
independent components) is taken by the metric induced on the 3-dimensional
hypersurfaces ($6$ variables) and their extrinsic curvature ($6$ variables),
which generate canonical pairs. Einstein equations are replaced by the
Hamiltonian evolution of these canonical pairs. Beside these there are
constraint equations to be fulfilled in each instant (on each hypersurface).
These are the Hamiltonian and Diffeomorphism constraints. In the generic
case of the 3+1 decompositions there is no preferred time, the formalism has
to be valid for any possible temporal choices ("many-fingered time"\
formalism \cite{manyfingeredtime, manyfingeredtime2}). 
Preferred choices 
arise by either imposing coordinate conditions \cite{K-coord1,K-coord2} or by
filling space-time with an adequate reference fluid \cite{K-ref1,K-ref2}.
Although the 3+1 decomposition breaks space-time covariance, 
a manifestly covariant canonical formalism based on the hyperspace 
(defined by all space-like hypersurfaces) has been also proposed 
\cite{K_hyp1,K_hyp2,K_hyp3,K_hyp4}.
The ADM decomposition
has been generalised for some of the modified gravity theories as well, e.g.
for the $f(R)$ gravitational theories \cite{Mongwane}.

If in addition a spatial direction plays a special role, the 2+1+1
decomposition of space-time may prove useful. 
This special role can be provided by a Killing-symmetry, for example 
the radial directions in either spherical or cylindrical symmetric space-times 
are such singled-out directions. We do not explore however simplifications 
arising from the imposition of symmetries, as for example applying 
mini-superspace or midi-superspace approaches \cite{K_cyl,K_SCHW}. 
The scenario we have in mind is to discuss generic perturbations of a 
background with certain symmetry.
In the most generic case both
singled-out directions have expansion, shear and vorticity \cite{Clarkson}.
The corresponding optical scalars were explored in the discussion of
perturbations of spherically symmetric space-times \cite{CB}, also for the
discussion of gravitational waves in anisotropic Kantowski-Sachs space-times 
\cite{KFBDG}.

In another, much simpler 2+1+1 decomposition formalism the decomposition is
made along a perpendicular double foliation \cite{s+1+1a, s+1+1b}. This
formalism has been employed in the framework of dark matter and dark energy
motivated scalar-tensor gravitational theories in the discussion of the odd
sector perturbations of spherically symmetric gravity in the effective field
theory approach \cite{KGT}. The requirement of perpendicularity however
consumes one gauge degree of freedom by fixing a metric function to vanish.
This has posed no problem in the discussion of the odd sector, however for
the even sector it generates an arbitrary function in the solution,
hampering the physical interpretation of perturbations. Therefore a modified
2+1+1 decomposition formalism would be desireble, which keeps the relative
simplicity of the formalism of \cite{s+1+1a, s+1+1b} (as compared to the
formalism exploring optical scalars \cite{Clarkson}), but employs $10$
metric functions instead of$\ 9$, hence becomes suitable for the discussion
of the even sector. Such a formalism could be worked out at the price of
relaxing the perpendicularity requirement \cite{2+1+1paper}.

In this conference report we summarise the main feature of this new
formalism and sketch the derivation of the Hamiltonian formalism, without
insisting on the involved computational details and related proofs of the
statements, which are given in Ref. \cite{2+1+1paper} together with
additional details.

Latin indices denote 4-dimensional space-time indices. Boldface lowercase
(as $\mathbf{i}$) or uppercase (as $\mathbf{A}$) latin letters count
2-dimensional or 4-dimensional basis vectors.

\section{The nonorthogonal double foliation}

\begin{figure}[H]
\begin{center}
\includegraphics[scale=0.55]{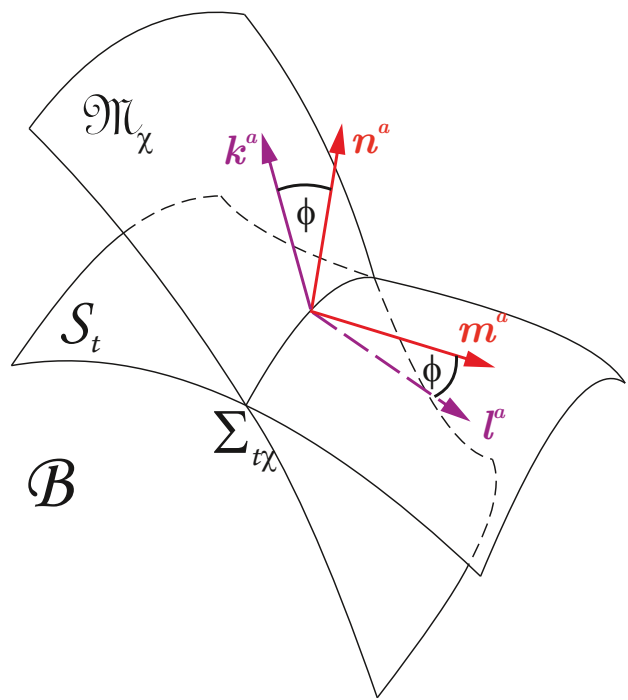}
\end{center}
\caption{The hypersurfaces of the nonorthogonal double foliation and the
adapted bases.}
\label{fig1}
\end{figure}

We generalise the orthogonal 2+1+1 decomposition of Refs. \cite{s+1+1a,
s+1+1b} such that the hypersurfaces $\mathcal{S}_{t}$ of constant $t$ and $\mathfrak{M}_{\chi}$ of constant $\chi$ with normal vectors $n^{a}$ and $l^{a}$, respectively, are nonorthogonal, as presented on Fig. \ref{fig1}. 

Their intersection is the surface $\Sigma _{t\chi }$, with an adapted vector
basis $\left\{F_{\mathbf{i}}\right\}$. We introduce two orthonormal bases
adapted to the two foliations, as follows: $f_{\mathbf{A}}=\left\{n,m,F_{\mathbf{i}}\right\}$ and $g_{\mathbf{A}}=\left\{k,l,F_{\mathbf{i}}\right\}$. The 4-dimensional metric can then be decomposed in both:
\begin{eqnarray}
\tilde{g}_{ab} &=&-n_{a}n_{b}+m_{a}m_{b}+g_{ab}~,  \label{gtildef} \\
\tilde{g}_{ab} &=&-k_{a}k_{b}+l_{a}l_{b}+g_{ab}~.  \label{gtildeg}
\end{eqnarray}
Here $g_{ab}$ is the metric induced on $\Sigma _{t\chi}$.

The temporal and selected spatial evolution vectors in the $f_{\mathbf{A}}$
basis are:
\begin{eqnarray}
\left( \frac{\partial }{\partial t}\right)^{a} &=&Nn^{a}+N^{a}+\mathcal{N}m^{a}~,  \label{ddt} \\
\left( \frac{\partial }{\partial \chi}\right)^{a} &=&Mm^{a}+M^{a}+\mathcal{M}n^{a}~.  \label{ddchi}
\end{eqnarray}
They define a coordinate-basis, the duality relations of which imply \cite{2+1+1paper} 
\begin{equation}
\mathcal{M}=0~,
\end{equation}
making manifest that $\partial /\partial \chi$ is tangent to $\mathcal{S}_{t}$.
The shift component $\mathcal{N}$ arises due to the nonorthogonality of the
foliations and generates all new terms arising as compared to the formalism
presented in Refs. \cite{s+1+1a, s+1+1b}, where $\mathcal{N}=0$ was imposed.
With the introduction of a nonvanishing $\mathcal{N}$ full gauge freedom is
reestablished, with $10$ metric components in the formalism 
($3$ for $g_{ab}$, $2$ for $M^{a}$ and $N^{a}$ each, one for each of the lapses $N$,~$M$ and
shift component $\mathcal{N}$). At times it will be convenient to
parametrize this $10^{th}$ metric function as $\mathcal{N}=N\tanh\phi$ and also 
employ the notations $\mathfrak{s}=\sinh\phi$, $\mathfrak{c}=\cosh\phi$. 
This is especially convenient in
proving \cite{2+1+1paper} that the two bases are related by a
Lorentz-rotation: 
\begin{equation}
\left( {\begin{array}{c}
   k^{a}\\
   l^{a}\\
  \end{array} } \right)
=
\left( {\begin{array}{cc}
   \mathfrak{c} & \mathfrak{s}\\
   \mathfrak{s} & \mathfrak{c}\\
  \end{array} } \right)
  \left( {\begin{array}{c}
   n^{a}\\
   m^{a}\\
  \end{array} } \right) 
\end{equation}
also to derive the decomposition of the evolution vectors in the basis $g_{\mathbf{A}}$:
\begin{eqnarray}
\left( \frac{\partial }{\partial t}\right)^{a} &=&\frac{N}{\mathfrak{c}}k^{a}+N^{a}~, \\
\left( \frac{\partial }{\partial \chi}\right)^{a} &=&M\left( -\mathfrak{s}k^{a}+\mathfrak{c}l^{a}\right) +M^{a}~.
\end{eqnarray}
Note that $\partial /\partial t$ is manifestly tangent to the hypersurface $\mathfrak{M}_{\chi}$.

Finally, in the basis $f_{\mathbf{A}}$\ it is straightforward to check 
\begin{equation}
\left[ m,F_{\mathbf{j}}\right] ^{a}n_{a}=0~,
\end{equation}
reassuring (due to the Frobenius Theorem) that $n^{a}$ is
hypersurface-orthogonal and 
\begin{equation}
\left[ n,F_{\mathbf{j}}\right] ^{a}m_{a}=\frac{M}{N}\partial _{\mathbf{j}}\left( \frac{\mathcal{N}}{M}\right) ~,
\end{equation}
implying that the vector $m^{a}$ has vorticity. Similarly, in the basis $g_{\mathbf{A}}$ we find that $l^{a}$ is hypersurface-orthogonal and $k^{a}$ has
vorticity:
\begin{eqnarray}
\left[ k,F_{\mathbf{j}}\right] ^{a}l_{a} &=&0~,  \label{kGi} \\
\left[ l,F_{\mathbf{j}}\right] ^{a}k_{a} &=&\frac{N}{\mathfrak{c}^{2}M}
\partial _{\mathbf{j}}\left( \frac{\mathfrak{sc}M}{N}\right) ~.
\end{eqnarray}
Hence the $10^{th}$ metric function $\mathcal{N}$ bears a double
interpretation: (1) it gives the angle of the Lorentz-rotation between the
two bases, and (2) generates the vorticity of the complementary basis
vectors $m^{a}$ and $k^{a}$. More details of these interpretations will be
presented in Ref. \cite{2+1+1paper}.

\section{The 2+1+1 decomposition of covariant derivatives}

The projected covariant derivative of any tensor $T_{b_{1}...b_{r}}^{a_{1}...a_{r}}$ defined on $\Sigma _{t\chi }$ arises by
projecting in all indices with $g_{a}^{b}$:
\begin{equation}
D_{a}T_{b_{1}...b_{q}}^{a_{1}...a_{r}}\equiv
g_{a}^{c}g_{c_{1}}^{a_{1}}...g_{c_{r}}^{a_{r}}g_{b_{1}}^{d_{1}}...g_{b_{q}}^{d_{q}}
\tilde{\nabla}_{c}T_{d_{1}...d_{q}}^{c_{1}...c_{r}}~.  \label{kovd}
\end{equation}
The $D$-derivative obtained in this way is related to the connection
compatible with the 2-metric due to the property
\begin{equation}
D_{a}g_{bc}=0~.
\end{equation}
It will be of particular importance to 2+1+1 decompose the covariant
derivatives of the basis vectors. We found:
\begin{equation}
\tilde{\nabla}_{a}n_{b}=K_{ab}+2m_{(a}\mathcal{K}_{b)}+m_{a}m_{b}\mathcal{K}-n_{a}\left( \mathfrak{a}_{b}-m_{b}\mathcal{L}^{\ast }\right) ~,
\label{nfelb}
\end{equation}
\begin{equation}
\tilde{\nabla}_{a}m_{b}=L_{ab}^{\ast }+n_{a}\mathcal{L}_{b}^{\ast }+n_{b}\mathcal{K}_{a}+n_{a}n_{b}\mathcal{L}^{\ast }+m_{a}\left( \mathfrak{b}_{b}^{\ast }+n_{b}\mathcal{K}\right) ~,  \label{mfelb}
\end{equation}
\begin{equation}
\tilde{\nabla}_{a}k_{b}=K_{ab}^{\ast }+l_{a}\mathcal{K}_{b}^{\ast }+l_{b}\mathcal{L}_{a}+l_{a}l_{b}\mathcal{K}^{\ast }-k_{a}\left( \mathfrak{a}_{b}^{\ast }-l_{b}\mathcal{L}\right) ~,  \label{kfelb}
\end{equation}
\begin{equation}
\tilde{\nabla}_{a}l_{b}=L_{ab}+2k_{(a}\mathcal{L}_{b)}+k_{a}k_{b}\mathcal{L}+l_{a}\left( \mathfrak{b}_{b}+k_{b}\mathcal{K}^{\ast }\right) ~,
\label{lfelb}
\end{equation}
where $K_{ab}=g_{a}^{c}g_{b}^{d}\tilde{\nabla}_{c}n_{d}$, $L_{ab}^{\ast
}=g_{a}^{c}g_{b}^{d}\tilde{\nabla}_{c}m_{d}$, $K_{ab}^{\ast
}=g_{a}^{c}g_{b}^{d}\tilde{\nabla}_{c}k_{d}$ and $L_{ab}=g_{a}^{c}g_{b}^{d}\tilde{\nabla}_{c}l_{d}$ are extrinsic curvatures of the surface $\Sigma
_{t\chi }$; $\mathcal{K}_{a}=g_{a}^{c}m^{d}\tilde{\nabla}_{c}n_{d}$ and $\mathcal{L}_{a}=-g_{a}^{c}k^{d}\tilde{\nabla}_{c}l_{d}$ are normal
fundamental forms; $\mathcal{K}=m^{d}m^{c}\tilde{\nabla}_{c}n_{d}$, $\mathcal{L}^{\ast }=n^{c}n^{d}\tilde{\nabla}_{c}m_{d}$, $\mathcal{K}^{\ast
}=l^{d}l^{c}\tilde{\nabla}_{c}k_{d}$ and $\mathcal{L}=k^{d}k^{c}\tilde{\nabla}_{c}l_{d}$ are normal fundamental scalars \cite{Schouten}. The quantities $\mathcal{L}_{a}^{\ast }=-g_{a}^{d}n^{c}\tilde{\nabla}_{c}m_{d}$ and $\mathcal{K}_{a}^{\ast }=g_{a}^{d}l^{c}\tilde{\nabla}_{c}k_{d}$ are defined
similarly to the normal fundamental forms, but they also contain the
contributions of the vorticities of the corresponding vectors. Finally $\mathfrak{a}_{a}=g_{a}^{d}n^{c}\tilde{\nabla}_{c}n_{d}$, $\mathfrak{b}_{a}^{\ast }=g_{a}^{i}m^{b}\tilde{\nabla}_{b}m_{i}$, $\mathfrak{a}_{a}^{\ast}=g_{a}^{i}k^{b}\tilde{\nabla}_{b}k_{i}$ and $\mathfrak{b}_{a}=g_{a}^{d}l^{c}
\tilde{\nabla}_{c}l_{d}$ are the projections onto $\Sigma _{t\chi }$ of the
nongravitational accelerations of the respective observers (among which
those moving along $n^{a}$ and $k^{a}$ are physical).

The set of above quantities is not independent. As shown in detail in Ref. 
\cite{2+1+1paper}, it is enough to select the sets $\left( K_{ab},~\mathcal{K}^{a},~\mathcal{K}\right)$, $\left( L_{ab}^{\ast },~\mathcal{L}^{\ast
}\right)$, $\left( \mathfrak{a}_{a},\mathfrak{b}_{a}^{\ast }\right)$ and $\mathcal{N}$\ in order to express all the others. In particular, for
orthogonal foliations all starry quantities reduce to nonstarred ones.
Beside, the set $\left( K_{ab},~\mathcal{K}^{a},~\mathcal{K}\right)$ is
related to time derivatives of the metric variables:
\begin{eqnarray}
K_{ab} &=&\frac{1}{N}\left[ \frac{1}{2}\partial _{t}g_{ab}-D_{(a}N_{b)}%
\right] -\frac{\mathfrak{s}}{M\mathfrak{c}}\left[ \frac{1}{2}\partial _{\chi
}g_{ab}-D_{(a}M_{b)}\right] ~, \\
\mathcal{K}^{a} &=&\frac{1}{2MN}\left( \partial _{t}M^{a}-\partial _{\chi
}N^{a}-N^{b}D_{b}M^{a}+M^{b}D_{b}N^{a}\right)  \\
&&-\frac{M}{2N}D^{a}\left( \frac{\mathcal{N}}{M}\right) ~,  \notag \\
\mathcal{K} &=&\frac{1}{MN}\left[ \partial _{t}M-\partial _{\chi }\mathcal{N}%
-N^{a}D_{a}M+M^{a}D_{a}\mathcal{N}\right] ~,
\end{eqnarray}
while the set $\left(L_{ab}^{\ast},~\mathcal{L}^{\ast }\right)$ is
connected to their $\chi $-derivatives only:
\begin{eqnarray}
L_{ab}^{\ast } &=&\frac{1}{M}\left[ \frac{1}{2}\partial _{\chi
}g_{ab}-D_{(a}M_{b)}\right] ~, \\
\mathcal{L}^{\ast } &=&-\frac{1}{M}\left[ \partial _{\chi }\left( \ln
N\right) -M^{a}D_{a}\left( \ln N\right) \right] ~.
\end{eqnarray}
Moreover, the accelerations can be expressed as $D$-derivatives of the
lapses:
\begin{eqnarray}
\mathfrak{a}_{b} &=&D_{b}\left( \ln N\right) ~, \\
\mathfrak{b}_{b}^{\ast } &=&-D_{b}\left( \ln M\right) ~.
\end{eqnarray}

\section{Hamiltonian dynamics}

The Einstein-Hilbert action
\begin{equation}
S_{EH}=\int d^{4}x\sqrt{-\tilde{g}}\tilde{R}  \label{S}
\end{equation}
can be rewritten by employing the twice contracted Gauss identity \cite{2+1+1paper} and the decomposition $\sqrt{-\tilde{g}}=NM\sqrt{g}$ as
\begin{eqnarray}
S_{EH} &=&S_{EH}\left[ \left\{ g_{ab},M^{a},M\right\} ;\left\{ K_{ab},%
\mathcal{K}^{a},\mathcal{K}\right\} ;\left\{ L_{ab}^{\ast },\mathcal{L}%
^{\ast }\right\} ;\left\{ N,N^{a},\mathcal{N}\right\} \right]   \notag \\
&=&\int dt\int d\chi \int_{\Sigma _{t\chi }}d^{2}xNM\sqrt{g}\left\{
R+K_{ab}K^{ab}-K^{2}-2K\mathcal{K}+2\mathcal{K}^{a}\mathcal{K}_{a}\right.  
\notag \\
&&-L_{ab}^{\ast }L^{\ast ab}+L^{\ast 2}-2\mathcal{L}^{\ast }L^{\ast
}+2\left( NM\right) ^{-1}D^{a}MD_{a}N  \notag \\
&&\left. -2\tilde{\nabla}_{a}\left[ \alpha ^{a}-\beta ^{\ast
a}-n^{a}K+m^{a}L^{\ast }\right] \right\} ,  \label{S2+1+1}
\end{eqnarray}
which beside scalars contains only tensors and vectors defined on $\Sigma
_{t\chi }$. The total covariant divergence is not yet decomposed, however
upon decomposition it will generate only boundary terms. The set of
variables $\left( g_{ab},M^{a},M\right) $ are the generalised coordinates, $\left( K_{ab},\mathcal{K}^{a},\mathcal{K}\right) $ the generalised
velocities, while $\left( L_{ab}^{\ast },\mathcal{L}^{\ast }\right) $ can be
perceived as shorthand notations for the $\chi $-derivatives of the
generalised coordinates. Similarly to the 3+1 decomposition, time
derivatives of $\left( N,~N^{a},~\mathcal{N}\right) $ do not emerge in the
action. The generalised momenta arise as derivatives with respect to the
time derivatives of the generalised coordinates as 
\begin{eqnarray}
\pi ^{ab} &=&\sqrt{g}M\left[ K^{ab}-g^{ab}\left( K+\mathcal{K}\right) \right]
~, \\
p_{a} &=&2\sqrt{g}\mathcal{K}_{a}~, \\
p &=&-2\sqrt{g}K~.
\end{eqnarray}

Then the action can be rewritted in an already Hamiltonian form as \cite{2+1+1paper}: 
\begin{eqnarray}
S_{EH} &=&\int dt\int d\chi \int_{\Sigma _{t\chi }}d^{2}x\left[ \pi ^{ab}\dot{g}_{ab}+p_{a}\dot{M}^{a}+p\dot{M}\right.  \notag \\
&&\left. -N\mathcal{H}_{\perp }^{G}-N^{a}\mathcal{H}_{a}^{G}-\mathcal{NH}_{\mathcal{N}}^{G}+Q\right]
\end{eqnarray}
where $Q$ is a sum of boundary terms, given explicitly in \cite{2+1+1paper},
while 
\begin{eqnarray}
\mathcal{H}_{\perp }^{G} &=&\sqrt{g}\left[ -M\left( R+3L^{\ast
ab}L_{ab}^{\ast }-L^{\ast 2}\right) +2g^{ab}\partial _{\chi }L_{ab}^{\ast
}\right.   \notag \\
&&-2\left( M^{c}D_{c}L^{\ast }+2L_{ab}^{\ast }D^{a}M^{b}\right) +2D^{a}D_{a}M
\notag \\
&&+\frac{M}{\sqrt{g}}\left[ \frac{1}{M^{2}}\left( \pi _{ab}\pi ^{ab}-\frac{%
\pi ^{2}}{2}\right) +\frac{1}{2}p_{a}p^{a}+\frac{1}{8}p^{2}-\frac{\pi p}{2M}%
\right] 
\end{eqnarray}
is the Hamiltonian constraint,
\begin{equation}
\mathcal{H}_{a}^{G}=-2D_{b}\pi _{~a}^{b}+pD_{a}M-\partial _{\chi
}p_{a}+p_{a}D_{b}M^{b}+M^{b}D_{b}p_{a}+p_{b}D_{a}M^{b}~,
\end{equation}
and
\begin{equation}
\mathcal{H}_{\mathcal{N}}^{G}=2L_{ab}^{\ast }\pi
^{ab}-2p^{a}D_{a}M-MD_{a}p^{a}-\partial _{\chi}p+D_{a}\left( pM^{a}\right)
\end{equation}
are the diffeomorphism contraints. Note that as expected, $\left( N,~N^{a},~\mathcal{N}\right)$ only appear as Lagrange-multipliers.

The evolution equations for the generalised coordinates and momenta then
emerge as the Hamiltonian equations written for the gravitational
Hamiltonian density 
\begin{equation}
\mathcal{H}^{G}=N\mathcal{H}_{\perp }^{G}+N^{a}\mathcal{H}_{a}^{G}+\mathcal{NH}_{\mathcal{N}}^{G}~.
\end{equation}
They are explicitly worked out in Ref. \cite{2+1+1paper}.

\section{Summary}

We generalised the formalism of \cite{s+1+1a,s+1+1b} by allowing for
nonorthogonal foliations. As main benefit, this led to the reestablisment of
the full gauge freedom, allowing a generic discussion of perturbations. We
gave a twofold geometrical interpretation the $10^{th}$ metric variable as
the angle of the Lorentz-rotation of the basis vectors and the measure of
the vorticity of the basis vectors.

In the ADM formalism the induced metric and extrinsic curvature of the
hypersurface play the role of Hamiltonian coordinates and momenta. In the
new formalism we identified those geometrical quantities characterising the
embedding, which bear dynamical role (they contain time derivatives).
Non-dynamical geometrical quantities appear only in the basis $f_{\mathbf{A}} $, hence we employed that for the 2+1+1 decomposition of the
Einstein-Hilbert action. From among the geometric variables we identified
those which combine into canonical pairs and proceeded with performing the
Hamiltonian analysis. We identified the 2+1+1 decomposed gravitational
Hamiltonian, also the Hamiltonian and momentum constraints in terms of
canonical coordinates and momenta.

We intend to apply this formalism both for the discussion of the even sector
of perturbations of spherically symmetric gravity in the effective field
theories of gravity and for the Hamiltonian treatment of canonically
quantisable cylindrical gravitational waves. The first of these has the
potential to address the stability of dark matter halo models in
scalar-tensor gravity. Also, for the discussion of gravitational waves in
space-times with particular symmetries, the 2+1+1 decomposition of the
Weyl-tensor would be an asset.

\vspace{6pt} 

\supplementary{The following are available online at www.mdpi.com/link, Figure S1: title, Table S1: title, Video S1: title.}

\acknowledgments{This work was supported by the Hungarian National Research Development
and Innovation Office (NKFI) in the form of the grant 123996. 
The work of C.G. was further supported by the UNKP-17-2 
New National Excellence Program of the Ministry of Human Capacities.
The work of Z.K. was further supported by the UNKP-17-4 New National Excellence Program of the Ministry of Human Capacities. 
C.G. and L.\'A.G. thank the organisers of the Bolyai-Gauss-Lobachevsky Conference for partial support of their participation.}

\authorcontributions{All authors contributed equally to this work.}

\conflictsofinterest{The authors declare no conflict of interest.} 



\reftitle{References}

\end{document}